\documentclass[journal=jpcc,manuscript=article,layout=twocolumn]{achemso}

\usepackage[version=3]{mhchem} 
\usepackage{color,soul}
\usepackage{stackengine}
\usepackage[]{caption}
\usepackage{graphicx}
\usepackage{subfig}
\usepackage{array}
\usepackage{soul,xcolor}                
\usepackage{comment}
\usepackage{mathtools}
\usepackage[hidelinks]{hyperref}

\newcommand*{\bA}{\ensuremath{\mathrm{A}}}  
\newcommand*{\bD}{\ensuremath{\mathrm{D}}}  

\newcommand*{\morsepot}{\ensuremath{U_{\ce{C-Pt}}}}
\newcommand*{\carbonheight}{\ensuremath{z_{\ce{C}}}}
\newcommand*{\carbonheightdag}{\ensuremath{z_{\ce{C}}^\ddag}}
\newcommand*{\probD}{\ensuremath{p_{\bD}}}
\newcommand*{\sigmaintrinsic}{\ensuremath{\sigma_{\mathrm{t}}}}
\newcommand*{\sigmahisto}{\ensuremath{\sigma}}
\newcommand*{\watercoord}{\ensuremath{\eta_{\mathrm{W}}}}
\newcommand*{\watercoorddag}{\ensuremath{\eta_{\mathrm{W}}^\ddag}}
\newcommand*{\anglecodag}{\ensuremath{\phi_{\ce{CO}}^\ddag}}
\newcommand*{\angleco}{\ensuremath{\phi_{\ce{CO}}}}
\newcommand*{\cutoff}{\ensuremath{r_\mathrm{c}}}
\newcommand*{\pathway}{\ensuremath{\chi}}
\newcommand*{\cdf}{\ensuremath{\mathcal{P}}}

\title[\\]{The influence of solvent on surface adsorption and desorption}

\author{Ardavan Farahvash}
\affiliation{Department of Chemistry, Massachusetts Institute of Technology, Cambridge, Massachusetts 02139, USA}
\altaffiliation{Contributed equally to this work}

\author{Mayank Agrawal}
\affiliation{School of Engineering, Brown University, Providence, Rhode Island 02912, USA}
\altaffiliation{Contributed equally to this work}

\author{Adam P. Willard}
\email{apwillard@mit.edu}
\affiliation{Department of Chemistry, Massachusetts Institute of Technology, Cambridge, Massachusetts 02139, USA}

\author{Andrew A. Peterson}
\email{andrew_peterson@brown.edu}
\affiliation{School of Engineering, Brown University, Providence, Rhode Island 02912, USA}

\begin{document}

\begin{abstract}
\noindent
    The adsorption and desorption of reactants and products from a solid surface is essential for achieving sustained surface chemical reactions.
    At a liquid--solid interface, these processes can involve the collective reorganization of interfacial solvent molecules in order to accommodate the adsorbing or desorbing species.
    Identifying the role of solvent in adsorption and desorption is important for advancing our understanding of surface chemical rates and mechanisms and for enabling the rational design and optimization of surface chemical systems.
    In this manuscript we use all-atom molecular dynamics simulation and transition path sampling to identify water's role in the desorption of CO from a Pt(100) surface in contact with liquid water.
    We demonstrate that the solvation of CO, as quantified by the water coordination number, is an essential component of the desorption reaction coordinate.
    We use meta dynamics to compute the desorption free energy surface and conclude based on its features that desorption proceeds via a two-step mechanism whereby the final detachment of CO from the surface is preceded by the formation of a nascent solvation shell.
\end{abstract}

\section{1. Introduction}
\label{sec:introduction}

    The reaction of atoms and molecules at solid surfaces is crucial to a myriad of technologies and processes.
    Much of our understanding of these processes comes from studies focused on the solid--vacuum interface, through low-pressure experimental and vacuum-surrounded atomistic studies; such studies are good proxies for reactions at the solid--gas interface\cite{jacobsen_catalyst_2001,norskov_towards_2009,norskov_density_2011,medford_sabatier_2015,liu_understanding_2017,wei_rational_2023,bullock_using_2020}.
    However, many increasingly important technologies and processes---including electrolyzers, batteries, fuel cells, corrosion, hydrothermal reforming, and some capacitors---take place at the solid--liquid interface, where comparably little is known about the influence of the liquid on surface dynamics \cite{franco_transition_2020,akbashev_electrocatalysis_2022,cortright_hydrogen_2002,huber_renewable_2004,gohda_influence_2009,carrasco_molecular_2012,liu_understanding_2017,xie_insights_2019,sakong_water_2020,ringe_double_2020}.
    The goal of this work is to use atomistic methods to explore the effect of the solvation on reactions at the solid--liquid interface.

    Adsorption and desorption are pivotal aspects of all surface-chemical processes.
    They represent both the initial and final steps of catalytic reactions, the charging/discharging mechanism in batteries, and they drive the process of corrosion.
    Adsorption reactions at the solid--gas interface have received decades of fundamental experimental and theoretical study.
    Experimental measurements have shown that the Arrhenius pre-exponential factor for desorption is anomalously high---often 10$^{14}$--10$^{17}$ s$^{-1}$, as opposed to ${\sim}10^{13}$ s$^{-1}$ as is more typical of reactions following transition-state theory.~\cite{Zhdanov1988,Chorkendorff2003,DellAngela2013}
    This high pre-factor is often attributed to a high-entropy transition state, since the desorbed state has a much higher entropy than the adsorbed state.
    Similar high pre-factors are seen in barrierless gas-phase dissociation reactions~\cite{Nazin1972} and are justified with similar reasoning.

    Computational tools, such as classical and first-principles molecular dynamics simulation, can be utilized to gain insight into the mechanisms of surface chemical processes.
    However, assessing the role of solvent with these tools presents significant practical challenges due to system size and sampling requirements.
    Simulation cells with dimensions larger than 1 nm$^3$ (and ideally much larger) are required to develop a solid-liquid interface that also includes a region of bulk liquid and an interfacial liquid surface that is free from finite-size effects (\textit{e.g.}, due to orientational correlations that span the periodic boundaries).
    In addition, these large simulations must be sampled in order to represent the effects of thermal fluctuations in local solvent structure and composition (\textit{e.g.}, due to the presence or absence of a dilute electrolyte species).
    Due to these challenges, molecular dynamics simulations of solvated reactions at metal interfaces often employ implicit solvent or small ice-like water layers. \cite{gohda_influence_2009,carrasco_molecular_2012,liu_understanding_2017,kastlunger_controlled-potential_2018} 
    These strategies provide valuable yet incomplete insight into reaction mechanisms.

    In this manuscript, we demonstrate how molecular dynamics can be utilized to determine the role of solvent in the adsorption/desorption of a small molecule from a solid surface.
    Specifically, we consider a model of CO desorption at the Pt(100)-liquid water interface.
    This system is both simple and ubiquitous---being an important step in many catalytic reactions and processes, such as carbon dioxide (\ce{CO_2}) electroreduction\cite{hori_electrochemical_2008}, Fischer-Tropsch synthesis \cite{van_der_laan_kinetics_1999}, automotive catalysts~\cite{Rood2019}, and supercritical-water biomass gasification~\cite{Antal2000}.
    
    The potential energy surface (PES) for CO desorption at a gas-phase Pt interface can be represented with only a few coordinates (\textit{e.g.}, CO-to-surface distance, CO orientation).
    In contrast, the PES for CO desorption at a liquid water-Pt interface is rugged and inherently high dimensional due to the influence of solvent configuration on the process.
    On this rugged landscape, there are numerous possible paths, corresponding to different molecular mechanisms.
    Determining which of these paths are relevant and which collective coordinates unite them can present a significant challenge.
    The method of transition path sampling (TPS), first proposed by Dellago et al.,\cite{dellago_efficient_1998,dellago_transition_1998} was developed to address this challenge.
    TPS combined with tools of statistical analysis and enhanced sampling can be used to identify and validate possible reaction coordinates that incorporate the collective solvent degrees of freedom, and derive reaction free energy surfaces based on these coordinates.

    In this study, we use TPS to identify a suitable reaction coordinate for CO desorption/adsorption that includes the collective solvent contribution.\cite{metad1}
    We determine that solvent reorganization, as quantified by the solvent coordination number, is an important component of the reaction coordinate, thereby demonstrating solvent's crucial role in the CO desorption pathway.
    Additionally, we show that the CO desorption pathway happens in a multi-step process, where the CO molecule first expands its solvation shell before leaving the surface. 
    By comparing results derived from the gas-phase and aqueous-phase systems, we illustrate how the presence of the liquid reshapes the desorption free energy profile.

\section{2. Methods}
\label{sec:methods}

\subsection{2.1 System details \label{subsec:sim_details}}
    
    \begin{figure}
        \centering
        {\includegraphics[width=1.0\columnwidth]{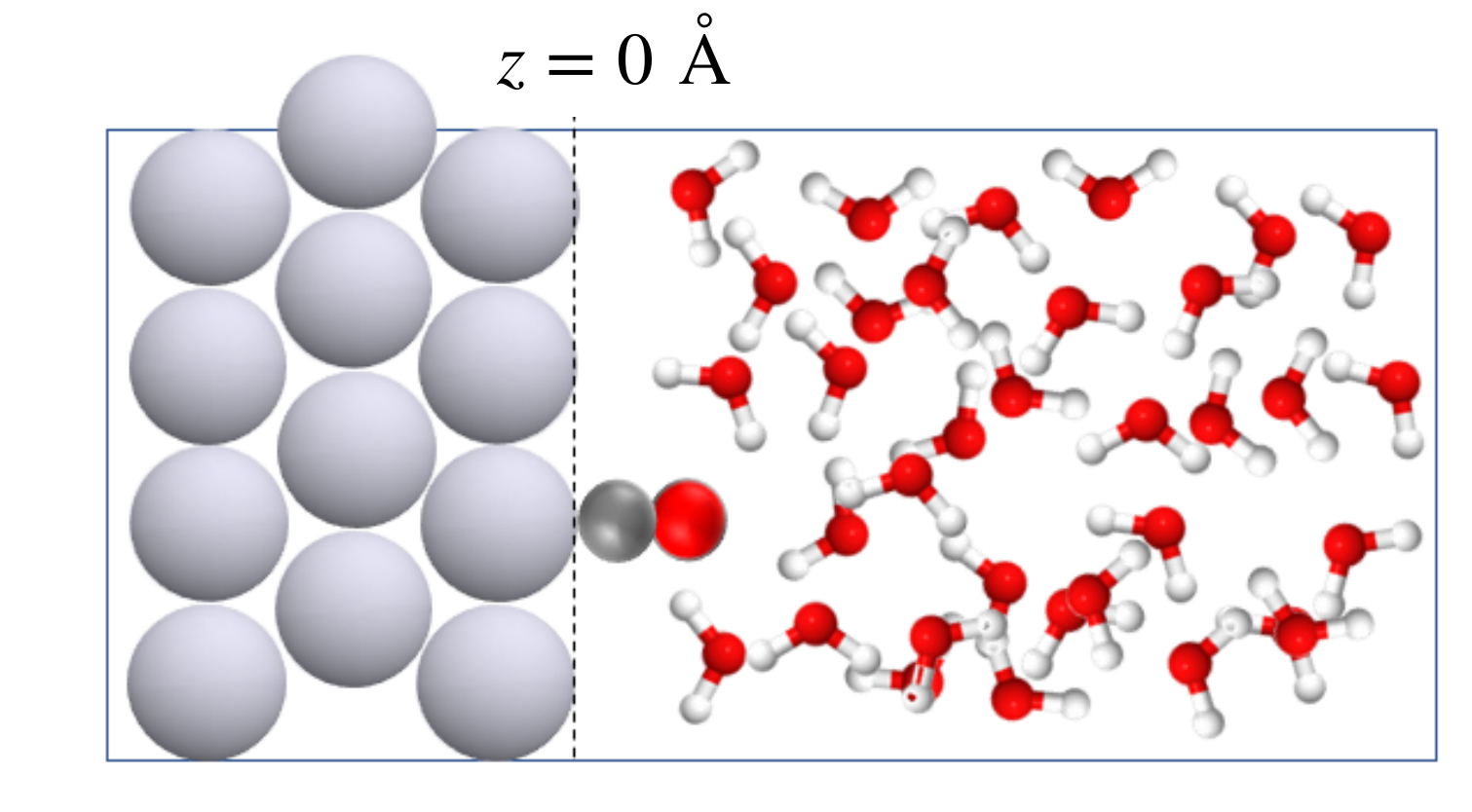}}
        \caption{Snapshot of simulation system consisting of a three-layer Pt(100) surface with CO molecule adsorbed atop a central site and 57 water molecules in 1650 \AA$^3$ volume above surface to achieve a water density of roughly 1 gm/cm$^3$.
        \label{fig:system}}
    \end{figure}
        
    Our basic model, illustrated in Figure \ref{fig:system}, consists of a  4$\times$4$\times$3 (atoms) Pt(100) surface with a single \ce{CO} molecule adsorbed atop a central surface site.
    The surface was solvated by a layer of 57 water molecules roughly 1.5~nm thick to yield a water density of 1 g/cm$^3$.
    The system was periodically replicated in the directions parallel to the Pt surface.
    Fixed, reflecting boundary conditions were used in the direction normal to the surface.
    A SPC-pol3 forcefield\cite{spcpol3} was used to model the water--water interactions.
    For metal--water interactions, we used universal force-field (UFF) parameters with Lorentz-Berthelot\cite{Lorentz1881} mixing rules.
    The metal and \ce{CO} also interact via UFF with Lorentz-Berthelot mixing rules.
    This potential prevents the oxygen atom and \ce{Pt} from colliding, but does not accurately capture the strength of the Pt--C interaction.
    To more accurately capture the interaction between the carbon atom of the CO molecule and the platinum surface, we fitted density functional theory calculations to a Morse potential,
    \begin{equation}
        \morsepot (\carbonheight) = D (1-e^{-\alpha(\carbonheight-z_0)^2})
    \end{equation}
    \noindent
    where \carbonheight\ is the displacement between the carbon atom of CO and and a Pt surface atom, the well-depth $D$ is 346 meV, the well width $\alpha$ is 3 \AA$^{-1}$, and the equilibrium C--Pt distance $z_0$ is 2.5 \AA. 
    These parameters were fit to DFT calculations of a single CO molecule adsorbed atop a Pt(100) site using BEEF-vdW functional. 
    A plane-wave basis set was used with a 600 eV energy cutoff. Electronic structure calculations were performed with the GPAW program \cite{gpaw1,gpaw2}. 
    All molecular-dynamics simulations were performed in LAMMPS~\cite{LAMMPS} using using a constant-temperature (NVT thermostat) at 300K, with Pt atoms held fixed.
    
    For TPS simulations, an additional Morse wall was set at $z_\mathrm{wall}=13.5$ \AA, with parameters $D=0.43$ eV, $\alpha=3$ \AA$^{-1}$, and $z_0=0.5$ \AA.
    The purpose of this additional potential is to introduce an artificial basin of attraction to the desorbed state, which otherwise presents a flat free energy profile.
    By introducing this basin of attraction, the TPS trajectories can be shorter in length, leading to improved computational efficiency.
    This basin is displaced from the transition state and thus does not influence the characteristics of the transition state.
    The basin is not present in the calculation of the free energy surfaces.

\subsection{2.1 Transition path sampling \label{subsec:tps}}
    
    Here we provide a brief overview of TPS theory and describe how we applied this theory to study CO desorption.
    Excellent reviews of TPS can be found in Refs.~\citenum{bolhuis_transition_2021,peters_reaction_2016,brotzakis_one-way_2016}. 
    TPS is a method for performing Monte-Carlo sampling of reactive trajectories.
    In this study, we define a reactive trajectory as one that connects an adsorbed state (\bA) and a desorbed state (\bD) within a specified time.

    The application of a TPS algorithm yields an ensemble of reactive trajectories.
    Each of these trajectories passes through a transition state, \textit{i.e.}, a nuclear configuration that is equally likely to evolve to visit the \bA\ or \bD\ state when given random momenta from a Maxwell--Boltzmann distribution.
    Indeed, the transition state ensemble (TSE) is formally defined as the set of nuclear configurations with a committor value of \probD\ = 0.5, were \probD\ is the probability that a configuration evolves to visit state \bD\ before it visits state \bA.
    We note that, in general, the value of \probD\ for a given configuration, and therefore the properties of the TSE, depend on how \bA\ and \bD\ are defined and on how trajectories are sampled during the implementation of TPS.

    Analysis of the TSE allows one to discern which collective coordinates are necessary to characterize transition states and are therefore good reaction coordinates.
    The statistics of committor values, quantified via the committor distribution function, $\cdf(\probD)$, can be analyzed to evaluate whether a proposed reaction coordinate is a good descriptor of the reaction mechanism.
    More specifically, the so-called \textit{histogram test} considers the shape of $\cdf(p_\bD)$ evaluated at the value of the reaction coordinate for which $\mu \equiv \langle \probD \rangle = 0.5$.\cite{peters_using_2006}
    A reaction coordinate that accurately captures the reaction mechanism will present a histogram ($\cdf(\probD))$ that is narrowly peaked at $\probD=0.5$.
    The width of the committor histogram is highly dependent on the size of the TSE and a large number of samples is often needed to converge statistics.
    Peters\cite{peters_using_2006} found that the standard deviation of the intrinsic committor distribution \sigmaintrinsic\ could be related to the standard-deviation of the committor histogram \sigmahisto\ by, 
    \begin{equation}
        \label{eq:peters}
        \sigmaintrinsic^2 = \frac{N}{N-1}\left( \sigmahisto^2 - \frac{1}{N} \mu(1-\mu)  \right).
    \end{equation}
    where $\mu$ is again the mean of the histogram, $\langle \probD \rangle$.
    Note that in the limit of a large number of samples in the TSE the \sigmahisto\ and \sigmaintrinsic\ variances become the same.
    Furthermore, as the shape of the histogram is highly dependent on the number of bins, it is common to fit the mean and standard-deviation to a beta-distribution model\cite{peters_ptpq_2010}.

    For this study, we defined the states \bA\ and \bD\ based on the distance of the CO carbon and a Pt surface atom.
    Specifically, we define the adsorbed state, \bA, as all configurations with $\carbonheight < 2.5$~\AA\ and the desorbed state, \bD, as all configurations with $\carbonheight > 5.0$~\AA.
    We used the open-source package \textit{openpathsampling} to generate 2000 different transition paths connecting the two basins.\cite{openps}
    We employed the spring algorithm \cite{brotzakis_one-way_2016} for our calculations, which is similar in concept the the shooting algorithm but randomizes the time index at which new trajectories are generated, picking a new random velocity at a given timestep. Note that spring shooting algorithms always use a stochastic thermostat. 
    Once a set of transition paths were generated, we extracted 1,000,000 randomly selected configurations residing between \bA\ and \bD\ to perform committer analysis.
    We assigned a configuration to the TSE if it had a committor value within a small range of the \probD\ = 0.5 configuration.
    The exact range of values are provided in Figures~\ref{fig:figure2}B. These values produced TSEs with roughly 1,000 configurations. 
            
\subsection{2.2 Meta-dynamics}
\label{subsec:metadynamics}

    After using TPS to identify a collective reaction coordinate, we utilized meta-dynamics to compute the associated free energy surface.
    Metadynamics, developed by Laio and Parrinello,\cite{metad1} samples rare events by adding an artificial Gaussian bias potential to the existing potential energy surface.
    The bias potential is updated at regular intervals in an MD simulation and pushes the system away from regions that have been heavily sampled.
    Meta-dynamics has been applied to numerous problems in chemistry and materials science \cite{metad_chem1, metad_chem2, metad_bio1, metad_bio2, metad_bio3,metad_matsci1,metad_matsci2}.
    We performed well-tempered meta-dynamics\cite{metad2, metad3} by adding a bias potential every 100 fs with a bias-factor of 5, Gaussian height 0.5 kcal/mol and width 0.2 in an NVT MD simulation at 300~K.
    We ran these MD simulations for 10 ns each to obtain a two-dimensional free-energy surface for the chosen collective coordinates through TPS; that is, the $z$-coordinate displacement between the carbon atom of the CO molecule and the nearest Pt surface atom (\carbonheight), and the water coordination around the CO molecule (\watercoord).
    Metadynamics simulations were performed using the PLUMED plugin in LAMMPS \cite{tribello_plumed_2014}. 
        
\section{3. Results and discussion}

\subsection{3.1 Relevant reaction coordinates for CO desorption}
    
    For any adsorption/desorption process, the direction normal to the metal surface (the $z$-axis in Figure~\ref{fig:system}) is an intuitive choice of reaction coordinate. 
    However, especially in the presence of a solvent, there may be other collective coordinates---for example the adsorbate's surroundings and its orientation---that are crucial to understand the transition states of the desorption process. 
    To determine the significance of a given collective variable, we use TPS and committor analysis, as described in the Section 2.1. 

    \begin{figure*}[h]
    \centering
    {\includegraphics[width=0.9\textwidth]{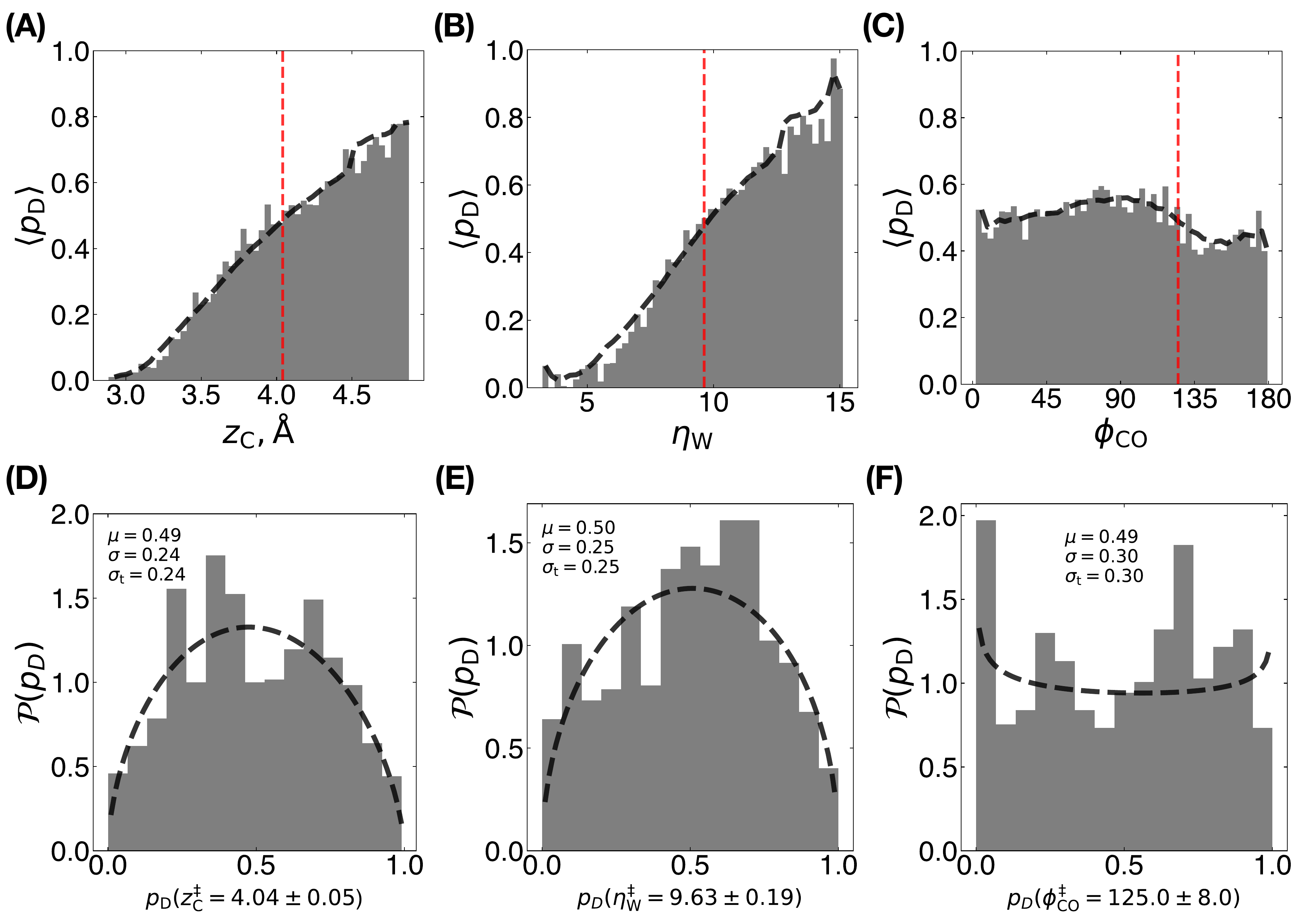}}
    \caption{
(A) Committor as a function of distance to surface.
The black dashed line is a windowed average over the bar heights with a window of 5.
The red vertical dashed line corresponds to the configuration closest to $\probD=0.5$.
(B) Committor as a function of water coordination number.
(C) Committor as a function of CO orientation.
(D) Committor histogram (gray bars) and beta distribution model (black dashed line) for configurations near \carbonheightdag\.
(E) Committor histogram and beta distribution model for configurations near \watercoorddag.
(F) Committor histogram and beta distribution model for configurations near \anglecodag.
    \label{fig:figure2}}
    \end{figure*}
        
    Figure \ref{fig:figure2}(A)--(C) shows the average committor probability \probD\ of a configuration reaching the desorbed state as a function of three collective coordinates: \carbonheight, \watercoord, and \angleco.
    \watercoord, which we henceforth term the \textit{solvent coordination number}, is defined as a function of the distance of the nearest neighboring water molecules to the CO molecule,
    \begin{equation}
        \label{eq:coord_num}
        \watercoord = 
        \sum_{i}
        \begin{dcases}
             0.5\left[ \cos \left( \frac{\pi \, d_i}{\cutoff} \right) + 1\right] & d_i < \cutoff \\[12pt]
            0 & d_i \geq \cutoff
        \end{dcases}
    \end{equation}
    where the index $i$ goes over all water molecules, $d_i$ is the distance between the carbon atom and the center of mass of water molecule $i$, and \cutoff\ is a cutoff radius, set to 6.5 \AA.
    Equation~\eqref{eq:coord_num} allows us to quantify the CO molecule's local solvation shell using a differentiable quantity suitable for the application of enhanced sampling methods.
    \watercoord\ is the CO molecule's orientation relative to the surface, with 0\textdegree\ corresponding to the carbon-down configuration. 

    As expected, the committor probability increased monotonically with \carbonheight, reinforcing the trivial notion that \carbonheight\ is an important collective variable. 
    We also observe that $\probD(\watercoord)$ increases monotonically, indicating that solvent coordination is also important.
    The Pearson correlation coefficient between \carbonheight\ and \watercoord\ taken over all configurations sampled is about 0.7, indicating that both collective coordinates, while correlated, contain some independent information. 
    In contrast, \angleco\ shows no clear correlation to the committor, indicating that it is not an important collective variable in determining the adsorption to desorption transition.
    We emphasize that this result does not indicate that \ce{CO} can freely adsorb in oxygen down configuration with no penalty. Instead, Fig~\ref{fig:figure2}C and ~\ref{fig:figure2}F illustrate only that the initial value of $ \angleco $ has no correlation with the final value after \ce{CO} either adsorbs or desorbs, indicating that there is no free-energy barrier along the $\phi_{\ce{CO}}$ coordinate.
    Previous experimental and theoretical studies have suggested that, in the gas phase, CO orientation can play an important factor in kinetics of adsorption and desorption when at sufficiently high temperature, due to the large loss in rotational entropy upon binding to the surface \cite{DellAngela2013,doren_precursor-mediated_1988,doren_dynamics_1991}.
    However, the same studies found that such effects were negligible at room temperature. 
    Since all our simulations were carried out at 300K, our finding that \angleco\ is not correlated with the committor is consistent with the existing literature for gas-phase desorption.
    However, we note that in general the committor is a temperature-dependent variable, and the conclusions here may not necessarily hold true at higher temperatures. 
    
    The red, vertical, dashed lines Figure \ref{fig:figure2}(A)--(C) represent the value of each coordinate for which the mean committor has the value $0.5$. 
    We henceforth denote the value of the collective coordinates at these transitions states with a superscript double dagger; \textit{e.g.}, \carbonheightdag.
    Figure \ref{fig:figure2}(D)--(F) presents the committor distribution function (gray bars) and corresponding beta-distribution estimates (black dashed lines), corresponding to roughly 1000 configurations taken near these transition states. 

    A good coordinate should display a histogram narrowly peaked around \probD\ = 0.5, meaning typical configurations at the nominal transition state also have \probD\ = 0.5.
    As expected, all of these distributions are have means at $\probD \approx 0.5$).
    Unsurprising the histogram/distribution for \angleco\ is particularly wide, suggesting that \anglecodag\ = 125\textdegree\ is not a good description of the transition state. 
            
    The committor distributions for both \carbonheightdag\ and \watercoorddag\ transition states are centered near the mean, however they are still are much wider than what is considered~\cite{peters_ptpq_2010} to be the gold-standard for a good collective coordinate ($\sigmaintrinsic=0.15$), illustrating that there are still many configurations with $\carbonheightdag \approx 4.04$ or $\watercoorddag \approx 9.63$ which do not necessarily have $\probD \approx 0.5$.
    Therefore, while both \carbonheight\ and \watercoord\ correlate strongly with adsorption/desorption process, alone they do not seem to capture all information necessary to capture the transition states, and therefore the kinetics. 

    \begin{figure}[h]
    \centering
    {\includegraphics[width=0.5\textwidth]{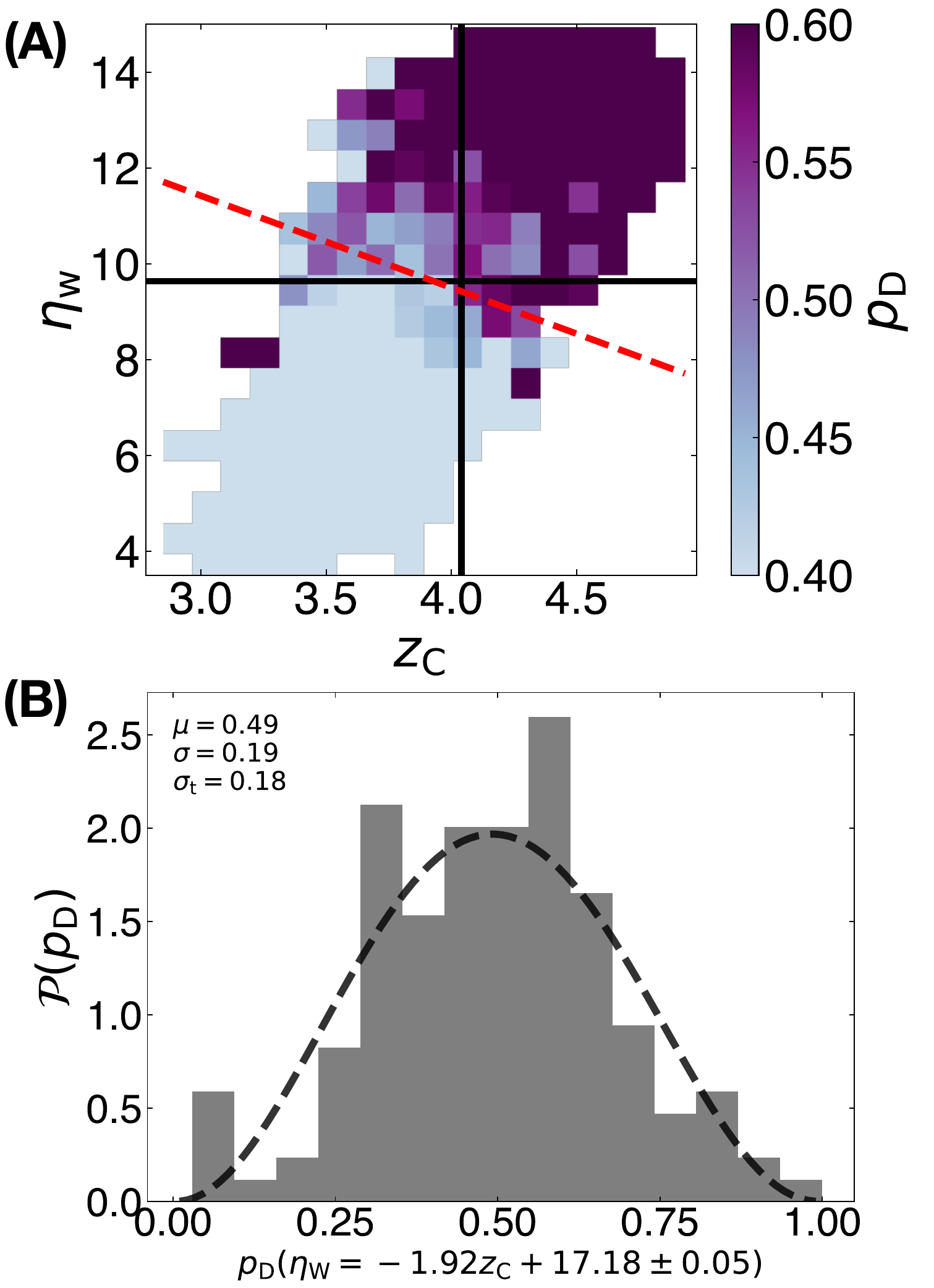}}
    \caption{
        (A) 2D committor probability as a function of the distance to surface and the water coordination number. The black vertical line corresponds to $z_\mathrm{C}=z_\mathrm{C}^\ddag$, the brown horizontal line corresponds to $\eta_\mathrm{W}=\eta_\mathrm{W}^\ddag$, and the red dashed line to states along  $\eta_\mathrm{W} = 1.92 z_\mathrm{C} + 17.18$.
        (B) Committor histogram and beta distribution model taken for states near the red dashed line in 3A. 
    \label{fig:figure3}}
    \end{figure}

    To further expand on these results, in Figure \ref{fig:figure3}(A) we present a 2-dimensional histogram of the committer probability as a function of \carbonheight\ and \watercoord\ simultaneously.
    The previous transition states \carbonheightdag\ and \watercoorddag\ are indicated in the black vertical and horizontal lines respectively.
    One can clearly see that many configurations along these lines have committor values either much higher or much lower than $\probD=0.5$.
    For example, a CO molecule at $\carbonheight = \carbonheightdag$ is very likely desorb if it also begins with a solvent coordination number $\watercoord > 12$.
    Instead of the one-dimensional transition-state surfaces defined by \carbonheightdag\ and \watercoorddag, a better transition-state surface can be defined along the line $\watercoord = 1.92 \carbonheight + 17.18$, shown as the red dashed line in  Figure~\ref{fig:figure3}(A).
    The committor histogram along this 2D transition-state surface, Figure \ref{fig:figure3}(B), has a significantly smaller standard deviation ($\sigmaintrinsic=0.18$) than considering $z_\mathrm{C}$ and $\eta_\mathrm{W}$ alone.
    While an improvement, we acknowledge that there may yet be some other subtle solvent coordinates playing a role in the desorption process. 
                    
\subsection{3.2 Free energy surface for CO desorption}

    \begin{figure*}
        \centering
        {\includegraphics[width=0.95\textwidth]{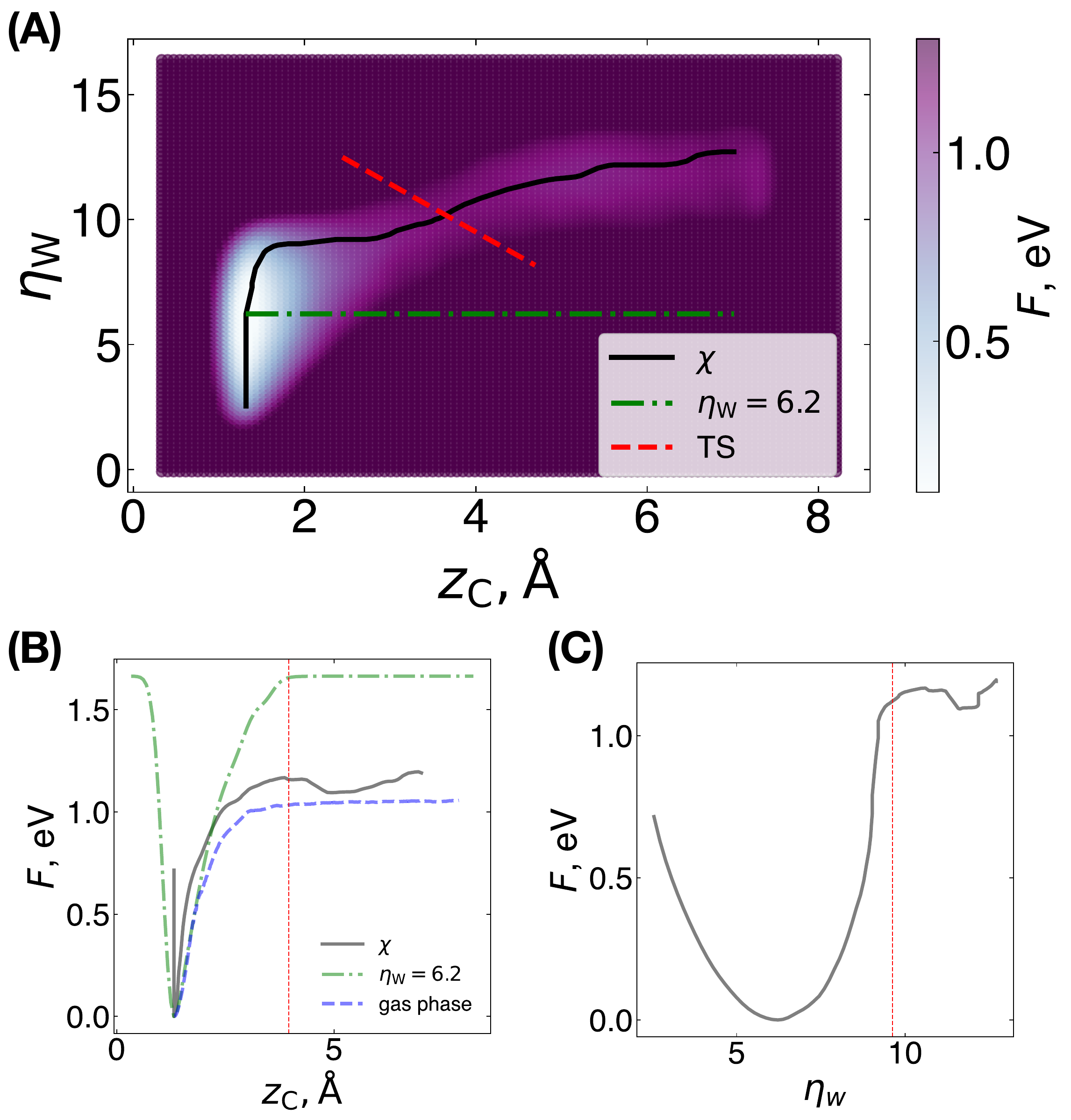}}
        \caption{
        (A) Free energy $F$ along \carbonheight\ and \watercoord\ calculated using metadynamics. The black contour corresponds to the path of minimum local free-energy stage (steepest descent) between reactants and products. A minor windowed average smoothing, with a window of 3 was applied to this contour to remove noise.
        The green contour represents a path of constant \watercoord.
        The red dashed line represents the transition state surface shown in Figure 2A found using TPS.
        (B) 1D cross-section of free-energy as a function of \carbonheight\ along the \pathway\ and $\watercoord=6.2$ contours, as well as the gas-phase desorption free-energy surface.
        (C) 1D cross-section of free-energy as a function of \watercoord\ along the \pathway\ contour.
        The red vertical lines show the $\probD=0.5$ states found with TPS. 
        \label{fig:figure4}}
    \end{figure*}

    In Figure~\ref{fig:figure4}, we present the 2-dimensional free-energy ($F$) surface, calculated with metadynamics, for CO desorption along the relevant coordinates identified by committor analysis, \carbonheight\ and \watercoord.
    Note that while the adsorption basin is clearly evident, with a minimum around $\carbonheight=1.3$ and $\watercoord=6.4$, the desorption basin is quite wide. 
            
    We considered two desorption pathways.
    The green line corresponds to a process where the water coordination number remains constant as the CO leaves the surface.
    The black line corresponds to the pathway along the gradient of the free-energy surface (henceforth denoted as \pathway) between the adsorbed and desorbed states.
    This is the path a CO molecule is most likely to travel.
    Analyzing the changes in slope of \pathway\ allows one to make conclusions about the steps and therefore mechanism of the desorption pathway.
    In particular, the results of Figure~\ref{fig:figure4}A suggest a multistep mechanism where, first, the solvent coordination number increases at roughly constant \carbonheight, second, the molecule increases its distance to the surface at roughly constant \watercoord, and third, both \carbonheight\ and \watercoord\ increase together. 
    The initial increase in the coordination number occurs due to a constriction of the first solvation shell around the CO molecule before it desorbs.
    
    In Figure~\ref{fig:figure4}B and Figure~\ref{fig:figure4}C, we present one-dimensional cross sections of $F$ for both the constant-\watercoord\ and \pathway\ pathways, and contrast them to the gas-phase Pt--CO potential of mean force for a single CO molecule desorbing in vacuum.  
    Comparing the dashed-dotted green line and solid black line reveals that the reaction barrier is much higher (by $\sim0.5$ eV) if CO does not increase it's coordination number while desorbing.
    Also notable is the presence of a barrier for the desorption of CO in solvent (along \pathway), which does not exist for the corresponding system in vacuum.
    This barrier/transition state is not a result of a single, high-energy, bottleneck conformation, as in gas phase reactions, but a collective effect arising from solvent reorganization.\cite{peters_common_2015}. 
    The collective nature of this transition state, similar to transition states observed in classical nucleation theory, supports why it is naturally broad.
            
    The red-dotted lines in Figure~\ref{fig:figure4} correspond to the transition states ($\probD=0.5$) we found using TPS.
    We see excellent agreement with the location of these lines and the location of the barriers on the free-energy surface.
    The agreement between TPS and metadynamics emphasizes the robustness of our results. 

    With regards to kinetics, the increased free-energy barrier to desorption between the gas and liquid phase ($\sim$0.2 eV), suggests a much slower desorption rate in solution.
    More interestingly, the fact that solvent reorganization must occur in order for the reagent to desorb highlights the importance of understanding surface--solvent interactions.
    A surface which interacts with solvent very strongly may hinder solute desorption not only directly, but through hindering solvent reorganization.
    In this study, we used a simple Lennard Jones model for the \ce{H2O}--Pt interaction, which leads to a somewhat weak interaction. We believe that a fruitful area for further research may be to study the effects of solvent-reorganization under different, and in particular stronger, surface-solvent potentials.

\section{4. Conclusions}
    In this paper, we have used enhanced sampling methods to analyze the role of solvent in the desorption of a model system of a CO molecule on a Pt(100) surface.
    We first employed transition path sampling to determine the relative importance of different collective coordinates. 
    Using the committor analysis we illustrated that both the distance to the surface and the solvent coordination number are needed to give a good description of transition states.
    Conversely, we found that the orientation of the CO molecule played a negligible role in the desorption process.
    We subsequently sampled the free-energy surface along our multi-dimensional reaction coordinate using metadynamics.
    Both methods showed excellent correspondence in the locations of transition states.
    Our metadynamics analysis indicated that desorption happens in a multi-step mechanism, where the initial step involves solvent reorganization and a nearly constant distance to the surface.
    Not allowing for solvent reorganization leads to a free-energy barrier that is significantly higher.
    Our results highlight the importance of the solvent for kinetics of reactions at interfaces.

\begin{acknowledgement}
This research was funded by the United States Department of Energy (DOE) under award DE‐SC0019441.
\end{acknowledgement}

%
%
%

\bibliography{refs_mayank.bib,refs_ardy}

\end{document}